\documentclass[a4paper,twoside]{article}

\usepackage{epsfig}
\usepackage{subfigure}
\usepackage{calc}
\usepackage{amssymb}
\usepackage{amstext}
\usepackage{amsmath}
\usepackage{amsthm}
\usepackage{multicol}
\usepackage{pslatex}
\usepackage{apalike}
\usepackage{SCITEPRESS}     
\usepackage{comment}
\usepackage{color}

\begin{document}

\title{The Curious Case of Machine Learning In Malware Detection}

\author{\authorname{Sherif Saad\sup{1}, William Briguglio \sup{1} and Haytham Elmiligi\sup{2}}
\affiliation{\sup{1}School Of Computer Science, Windsor University, Canada}
\affiliation{\sup{2}Computing Science Department, Thompson Rivers University ,Canada}
\email{{shsaad, briguglwb}@uwindsor.ca, helmiligi@tru.ca}
}


\abstract{In this paper, we argue that detecting malware attacks in the wild is a unique challenge for machine learning techniques. Given the current trend in malware development and the increase of unconventional malware attacks, we expect that dynamic malware analysis is the future for antimalware detection and prevention systems.  A comprehensive review of machine learning for malware detection is presented. Then, we discuss how malware detection in the wild present unique challenges for the current state-of-the-art machine learning techniques.  We defined three critical problems that limit the success of malware detectors powered by machine learning in the wild. Next, we discuss possible solutions to these challenges and present the requirements of next-generation malware detection.  Finally, we outline potential research directions in machine learning for malware detection.}

\onecolumn \maketitle \normalsize \vfill

\section{\uppercase{Introduction}}
\label{sec:introduction}

\noindent Nowadays, computer networks and the Internet have become the primary tool for spreading and distributing malware by malware authors. The massive number of \textcolor{black}{feature-rich} programming languages and off-the-shelf software libraries enable the development of new sophisticated malware such as botnet, fileless, k-ary and ransomware. New computing paradigms, such as cloud computing and the Internet of Things, expand potential malware infection sites from PC's to any electronic device.  


 To decide if software code is malicious or benign we could either use static analysis or dynamic analysis. Static analysis techniques do not execute the code and only examine the code structure and other binary data properties. Dynamic analysis techniques, on the other hand, execute the code to observe the execution behaviors of the code over the network or at endpoint devices. Some malware detection systems apply only static or dynamic techniques, and some apply both. While dynamic malware analysis techniques are not intended to replace static analysis techniques, recent unconventional malware attacks (botnet, ransomware, fileless, etc) and the use of sophisticated evasion techniques to avoid detection have shown the urgent need of dynamic analysis and the limitations of static analysis. In our opinion, the use of dynamic and behavioral malware analysis will dominate the next-generation malware detection systems. 


There is a general belief among cybersecurity experts that antimalware tools and systems powered by artificial intelligence and machine learning will be the solution to modern malware attacks.  The number of studies published in the last few years on malware detection techniques that leverage machine learning is \textcolor{black}{a} distinct evidence of this belief as shown in section \ref{sec:review}. In the literature, various malware detection techniques using machine learning are proposed with excellent detection accuracy. However, malware attacks in the wild continue to grow and manage to bypass malware detection systems powered by machine­learning techniques. This because it is difficult to operate and deploy machine learning for malware detection in a production environment or the performance in a production environment is disturbing (e.g., high false positives rate). In fact, there is a significant difference  (a detection gap) between the accuracy of malware detection techniques in the literature and their accuracy in a production environment.

A perfect malware detection system will detect all types of malicious software and will never consider a benign software as a malicious one. Cohen provided a formal proof that creating a perfect malware detection system is not possible \cite{cohen87,cohen89}.  Moreover,  Chess and White proved that a malware detector with zero false positives is not possible \cite{chess01}.  Selcuk et al. discussed the undecidable problems in malware detection in more details \cite{Selcuk17}. In light of this, the high levels of accuracy claimed by commercial malware detection systems and some malware detection studies in literature seems questionable.

In this paper, we briefly review the current state of the art in malware detection using machine learning approach. Then, we discuss the importance of dynamic and behavioral analysis based on emerging malware threats. Next, the shortcomings of the current machine learning malware detectors are explained to indicate their limitations in the wild. Finally, we discuss the possible solutions to improve the quality of malware detection systems and point out potential research directions.

\section{\uppercase{Literature Review}}
\label{sec:review}

In recent years, machine learning algorithms have been used to design both static and dynamic analysis techniques for malware detection.  Hassen et al. proposed a new technique for malware classification using static analysis based on control statement shingling \cite{Hassen17}. In their work, they used static analysis to classify malware instance into new or known malware families. They extracted features from disassembled malicious binaries and used random forest algorithm to classify malware using the extracted features. Using a dataset of 10,260 malware instances, they reported up to 99.21\% accuracy. 

Static analysis has been used to study malware\textcolor{black}{s} that infect embedded systems, mobile devices, and other IoT devices. Naeem et al. proposed a static analysis technique to detect IoT malware \cite{Naeem18}. The proposed technique converts a malware file to a grayscale image and extracts a set of visual features from the malware image to train an SVM classifier that could distinguish between malware families using visual features.  Using a dataset of 9342 samples that belong to 25 malware families, they reported 97.4\% accuracy. \textcolor{black}{Su et al.} proposed a similar technique to classify IoT malware into malware families using visual features and image recognition \cite{Su18}. Their approach is very similar to the one proposed in \cite{Naeem18}. They used a one-class SVM classifier and tested their approach on IoT malware\textcolor{black}{s} that infect Linux-like IoT systems; they reported 94.0\% accuracy for detecting malware and 81.8\% accuracy for detecting malware families. \textcolor{black}{Raff et al.} proposed a malware detection technique using static analysis and deep learning \cite{Raff17}. The proposed technique achieved 94.0\% detection accuracy. 

Several works have been proposed to detect Android malware apps using static analysis techniques.  Sahin et al. proposed an Android malware detection model that uses app permission to detect malicious apps \cite{Sahin18}. They used the permissions required by the app with a weighted distance function and KNN and Naive Bayes classifier to detect malicious apps.  They reported an accuracy up to 93.27\%. Su and Fung used sensitive functions and app permissions to detect Android malware \cite{YangSu16}. They used different machine learning algorithms such as SVM, decision tree, and KNN to build an android malware detector. They reported an average accuracy between 85.0\% and 90.0\%

Collecting and monitoring all malware behaviors is a complicated and time-consuming process. For that reason, several works in the literature focused on collecting partial dynamic behaviors of the malware. Lim et al. \cite{Lim15} proposed a malware detection technique by analyzing network traffic generated when the malware communicates with a malicious C\&C server such as in the case of botnet or ransomware. The proposed technique extracts a set of features from network flows to present a flows sequence. The authors used different sequence alignment algorithms to classify malware traffic. They reported an accuracy above 60\% when analyzing malware traffic in a real network environment.  

Kilgallon et al. applied machine learning and dynamic malware analysis \cite{Kilgallon17}. The proposed technique gathers register value information and API calls made by the monitored malware binaries. The collected information is stored in vector structures and analyzed using a value set analysis method. Then\textcolor{black}{,} they used a linear similarity metric to compare unseen malware to known malware binaries. Their experiment showed that the proposed technique could detect malware with an accuracy up to 98.0\%

Omind and Nathan proposed a behavioral-based malware detection method using a deep belief network \cite{ODavid15}.  The proposed method collected data about malware behaviors from a sandbox environment. The collected data is API calls, registry entries, visited websites, accessed ports, and IP addresses.  Then using a deep neural network of eight layers, it generates malware signatures.  These signatures could be used to train malware detectors. In their experiments, they reported up to 95.3\% detection accuracy with a malware detector utilizing the SVM algorithm.

Yeo et al.  proposed a new malware detection method by monitoring malicious behaviors in network traffic \cite{MYeo18}. They designed 35 features to describe malicious traffic of malware instances. They tested several machine learning algorithms including CNN, MLP, SVM, and random forest. The proposed method achieved an accuracy above 85\% when utilizing CNN or random forest. Prokofiev et al. proposed a machine-learning technique to detect C\&C traffic of infected IoT devices \cite{Prokofiev18}. The proposed approach used network traffic features such as port number, IP addresses, connection duration and frequency. They reported a detection accuracy up to 97.3\%. However, the proposed approach is still relying on traditional malware analysis methods and will not be able to work in production IoT deployment as discussed in \cite{Soliman17}. 
Several hybrid malware detection techniques that combine both static and dynamic analysis have also been proposed \cite{Martinelli16,DePaola18}. These techniques try to improve the quality and performance of malware detection systems by taking advantage of static and dynamic analysis to build robust malware detection systems.\\

\section{\uppercase{Emerging Malware Threats}}
\label{sec:emerging malware}
\noindent With the recent changes in malware development and the rise of commercial malware (malicious code rented or purchased), many new challenges are facing malware analysts that make static analysis more difficult and impractical. These challenges will force antimalware vendors to adapt behavioral malware analysis and detection techniques. In our opinion, \textcolor{black}{there are} two main reasons behind these challenges; the rise of unconventional computing paradigms and unconventional evasion techniques. There is a new generation of malware\textcolor{black}{s} that take advantage of unconventional computing paradigms and off-the-shelf software libraries written by feature-rich programming languages. The current state-of-the-art malware analysis/detection techniques and tools are not effective against this new generation of malware.

\subsection{Unconventional Computing Paradigms}
\noindent New computing paradigms and technologies such as cloud computing, the internet of things, big data,  in-memory computing, and blockchain introduced new playgrounds for malware authors to develop complex and sophisticated malware\textcolor{black}{s} that \textcolor{black}{are} almost undetectable. Here we describe several recent examples of new malware threats that are difficult to detect or analyze using static analysis. 

For instance\textcolor{black}{,} the Internet of Things (IoT) is an appealing platform for modern and sophisticated malware scuh as ransomware. \textcolor{black}{Zhang-Kennedy et al.}  discussed the ransomware threat in IoT and how a self-spreading ransomware could infect an IoT ecosystem \cite{Kennedy18}. The authors pointed out that the ransomware will mainly lock down IoT devices and disable the essential functions of these devices. The study focused on identifying the attack vectors in IoT, the techniques for ransomware self-spreading in IoT, and predicting the most likely class of IoT applications to be a target for ransomware attacks.  Finally, the authors identified the techniques the ransomware could apply to lock down IoT devices. \textcolor{black}{Authors in \cite{Kennedy18} used a Raspberry to develop a proof of concept IoT ransomware that can infect an IoT system}. One interesting aspect in \cite{Kennedy18} is the need for collaboration or swarming behavior in IoT ransomware, where the IoT ransomware will spread as much as possible and then lock down the devices or lock down the device and then spread.

\textcolor{black}{Miller and Valasek} developed a proof-of-concept for malicious code that infects connected cars and lockdown key functions \cite{Miller15}. For instance, the authors demonstrated the ability for the malicious code to control the steering wheel of a vehicle, disable the breaks, lock doors, and shut down the engine while in motion. Behaving as ransomware, this real example of \textcolor{black}{a} malware that locks and disables key features in IoT systems (e.g. connected cars) could have life threatening consequences if the ransom is not paid. The study explained a design flow in the Controller Area Network (CAN) protocol that allows malicious and crafted CAN message\textcolor{black}{s} to be injected into the vehicle CAN channel by a compromised mobile phone that is connected to the vehicle entertainment unit. It was reported that for some vehicles only the dealership could restore and patch the vehicle to prevent this attack. \textcolor{black}{Choi et al.} proposed a solution for malware attacks in connected vehicles using machine learning \cite{Choi18}. The solution uses SVM to distinguish between crafted malicious CAN messages, and benign CAN messages generated by actual electronic control units (ECU). The model extracts features from the vehicle ECUs and creates fingerprints for those ECUs.  The ECU fingerprint is noticeable in a benign CAN message and does not exist in a malicious message

\textcolor{black}{Azmoodeh et al.} discussed a new technique to detect ransomware attacks in IoT systems by monitoring the energy consumption of infected devices \cite{Azmoodeh2018}. As a proof of concept, they studied the energy consumption of infected Android devices. The devices were infected by a ransomware with crypto impact. They used different machine learning models (KNN, SVM, NN, and Random Forest) to analyze energy consumption data and extract unique patterns to detect compromised Android devices.  They reported a ransomware detection accuracy of 95.65\%. 

In 2015, Karam (INTERPOL) and Kamluk (Kaspersky lab) introduced a proof of concept distributed malware that also takes advantage of blockchain technology \cite{Karam11}. In 2018, Moubarak and et al. provided design and implementation of a K-ary malware (distributed malware) that takes advantages of the blockchain networks such as Etherum and Hyperledger \cite{Moubarak18}. The proposed malware is stored and executed inside blockchain networks and acts as a malicious keylogger. While detecting a K-ary malware is an NP-hard problem\cite{deDr06}, it is also complicated to implement a K-ary malware. However, Mubarak\textcolor{black}{'s} works demonstrated the simplicity of K-ary malware development by taking advantage of blockchain technology as distributed and decentralized network.

\subsection{Unconventional Evasion Techniques}
\noindent The new generation of malware will use advanced evasion techniques to avoid detection by antimalware systems and tools.  New evasion techniques implemented by malware authors use new technologies and off-the-shelf software libraries that enable the design of sophisticated evasion methods. Antimalware vendors and malware researchers discussed recent examples of using new antimalware evasion techniques in the wild.

Fileless malware or memory-resident malware is the new technique used by malware authors to develop and execute malicious attacks.  Fileless malware \textcolor{black}{resides} in device memory and does not leave any files on the infected device file system. This makes the detection of the fileless malware using signature-based detection or static analysis infeasible.  
In addition, the fileless malware takes advantage of the utilities and libraries that already exist in the platform of the infected device to complete its malicious intents. In other words, benign applications and software libraries are manipulated by fileless malware to accomplish the attack objectives.

Fileless malware attacks and incidents are already observed in the wild compromising large enterprises.  According to KASPERSKY lab, 140 enterprises were attacked in 2017 using fileless malware\textcolor{black}{s} \cite{KASPLab17}. Ponemon Institute reported that 77\% of the attacks against companies use fileless techniques \cite{Barkly17}. 
Moreover, there are several signs that ransomware attacks are going fileless, as discussed in \cite{Cyren18}. Besides these signs, there are other reasons in our opinion that \textcolor{black}{confirms that} ransomware and other malware attacks will be fileless. One main reason is the moving towards in-memory computing.

In recent years\textcolor{red}{,} in-memory computing and in-memory data stores became the first backbone and storage technology for many organizations. Many big data platforms and data grids (Apache Spark, Redis, HazelCast, etc.) enable storing data in memory for performance and scalability requirements.  Valuable data and information is stored in memory for a long time before moving to a persistent data store.  In-Memory ransomware that encrypts in-memory data (such as recent transactions,  financial information, etc.) present a severe and aggressive attack. \textcolor{black}{This is because} any attempt to reset or report the machine to remove the ransomware from the device memory or shutdown the application will result in losing this valuable data permanently.   

The moving towards distributed and decentralized computing is another reason for the rise of fileless ransomware. In distributed and decentralized computing several nodes and devices are available to store the in-memory malware, which will increase the life expectancy of the malware since there will always be a group of active node\textcolor{black}{s} were the malware could replicate and store itself.

The recent and massive development in machine learning /artificial intelligence (aka data science) and a large number of off-the-shelf machine learning libraries enable malware authors to develop advanced evasion techniques.  

\textcolor{black}{Rigaki and Garcia} proposed the use of deep learning techniques to create malicious malware samples that evade detection by mimicking the behaviors of benign applications \cite{Rigaki18}. In \textcolor{black}{their} work, \textcolor{black}{a} proof of concept was proposed to demonstrate how malware authors could cover the malware C\&C traffic.  The authors use a Generative Adversarial Networks (GANs) to enable malware (e.g., botnet) to mimic the traffic of a legitimate application and avoid detection. The study showed that it is possible to modify the source code of malware to receive parameters from a GAN to change the behaviors of its C\&C traffic to mimic the behaviors of other legitimate network applications\textcolor{black}{,} such as Facebook traffic.  The enhanced malware samples were tested against the Stratosphere Linux IPS (slips) system, which uses machine learning to detect malicious traffic.  The experiment showed that 63.42\% of the malicious traffic was able to bypass the detection.


A research team from IBM demonstrated the use of artificial intelligence to engineering malware attacks \cite{kirat18}. In \textcolor{black}{their} study, the author\textcolor{black}{s} proposed DeepLocker as a proof of concept to show how next-generation malware could leverage artificial intelligence. DeepLocker is a malware generation engine that malware author could use to empower traditional malware samples such as WannaCry with artificial intelligence. A deep convolutional neural network (CNN) was used to customize a malware attack by combining a benign application and a malware sample to generate a hybrid malware that bypasses detection by exposing (mimicking) benign behaviors. Besides that, the malware is engineered to unlock its malicious payload when it reaches a target (endpoint) with a loose predefined set of attributes. In the study, those attributes were the biometrics feature of the target such as facial and voice features. The malware uses CNN to detect and confirm target identity, and upon target confirmation, an encryption key is generated and used by the WannCry malware to encrypt the files on the target endpoint device. The encryption key is only generated by matching the voice and the facial features of the target. This means reverse engineering the malware using static analysis is not useful to recover the encryption key.

\section{\uppercase{Practical Challenges}}
 \label{sec:challenges}
 \noindent The new and emerging malware threats discussed in section \ref{sec:emerging malware} provide strong evidence for the need of adopting dynamic and behavior\textcolor{black}{al} analysis to build malware detection tools. The use of machine learning is the most promising technique to implement malware detectors and tools that apply behavior\textcolor{black}{al} analysis as shown in section \ref{sec:review}.  While the use of machine learning for malware detection has shown promising results in both static and dynamic analysis, there are significant challenges that limit the success of machine learning based malware detectors in the wild.  

\subsection{Cost of \textcolor{black}{Training Detectors}}  
\noindent The first challenge is the cost of training and updating malware detectors in production environment. Malware detection is unlike other domains where machine learning techniques have been applied successfully such as computer vision, natural language processing, and e-commerce. Malware instances evolve and change their behaviors over a short period; some studies by antimalware vendors reported that a new malware instance could change its behaviors in less than 24 hours since it has been released \cite{Gupta09,Allix15}.  This means a frequently trained machine learning model will become outdated. This also mean\textcolor{black}{s} we need to frequently retrain our malware detectors to be able to detect new and mutated malware instances. Therefore\textcolor{black}{,} adaptability in machine learning models for malware detection is a crucial requirement and not just a ancillary capability.  

Recently\textcolor{black}{,} the challenge of adaptability, and scalability of machine learning models for malware detection in the wild has become obvious \cite{Narayanan16}. The majority of the work proposed in the literature have done very little to reduce and optimize the feature space to design detectors ready for early malware detection in a production environment \cite{Hajmasan17}.  For instance, it is not clear how the proposed detection methods will scale when the number of monitored endpoints increase. Unlike computer vision, natural language processing and other areas that utilize machine learning, malware instances continue to evolve and change. This mostly requires retraining machine learning models in production, which is an expensive and complicated task. Therefore, when using machine learning for malware detection, we need to think differently. New methods to reduce the cost of retraining malware detectors and improve the detection quality are urgent.


\subsection{Malware Detector Interpretability}
Cybersecurity analysts always prefer solutions that are interpretable and understandable, such as rule-based or signature-based detection.  This is because of the need to tune and optimize these solutions to mitigate and control the effect of false positives and false negatives.  Interpreting machine learning models is a new and open challenge \cite{SShirataki17}. However, it is expected that an interpretable machine learning solution will be domain specific, for instance, interpretable solutions for machine learning models in healthcare are different than solutions in malware detection \cite{MAHMED18}. 

 Any malware detector will generate false positives, and unless malware analysts can understand and interpret the reason that a benign application wrongly classified as malicious, they will not accept those black box malware detectors. To our knowledge, no work in the literature investigated the interpretability of machine learning models for malware detection.

\subsection{Adversarial Malware}
Last but not least, a  malware detection system utilizing machine learning could be defeated (bypassed) using adversarial malware samples. For instance, Kolosnjaji et al. showed in \cite{Kolosnjaji18} that by using an intelligent evasion attack they can defeat the deep learning detection system proposed in \cite{Raff17} by Raff et al. They simply used their knowledge of how the proposed deep learning detection system operates and designed a gradient-based attack as an evasion technique to overcome it. With adversarial malware, the system detection accuracy dropped from 94.0\% to almost 50.0\%.  Machine learning algorithms are not designed to work with adversarial examples.  Grosse et al. demonstrated that using adversarial malware samples; they could reduce the detection accuracy of a malware detection system that uses static analysis and machine learning to 63.0\% \cite{Grosse17}. They also showed that adopting anti adversarial machine learning techniques used in computer vision is not effective in malware detection. Yang et al. proposed adversarial training as a solution for adversarial malware \cite{Yang2017MalwareDI}. They designed a method for adversarial android malware instances generation. The proposed method requires access to the malware binaries and source code, besides, it is mainly useful for static malware detection systems.  

\section{\uppercase{Bridging the Detection Gap}}
\noindent To overcome the challenges we discussed in section \ref{sec:challenges}, we propose \textcolor{black}{new solutions to} mitigate these challenges and reduce the gap.

\subsection{Disposable Micro Detectors} 
Current best practice\textcolor{black}{s} in constructing and building machine learning models follow a monolithic architecture. In monolithic architecture\textcolor{black}{,} a computationally-expensive single-monolithic (to build and train) machine learning model is used to detect malware\textcolor{black}{s}. While this architecture or approach for building machine learning models is successful in other domains, we \textcolor{black}{believe} it is unsuitable for malware detection given the highly evolving characteristic\textcolor{black}{s} of malware instance. We propose a new approach inspired by microservices architecture.  In this approach, multiple, small, inexpensive, focused machine learning models are built and orchestrated to detect malware instances. Each model or detector is built to detect the behaviors of a specific malware instance (e.g., Mirai, WannaCry), or at most a single malware family (a group of similar malware instances). Also, each model or detector is built using features that are similar, such as having the same computational cost, or unique to the specific execution environment. This is because out of the superset of features designed to detect malware, it is common that a subset of these features could be more or less useful to detect a specific malware instance or family.  
The use of micro (small) and focused detectors reduce the cost of retraining and deployment in production. This is because detectors for new malware could be trained and added without the need to retrain existing detectors. In addition, when a malware detector becames outdated as a result of malware evolving behaviors\textcolor{black}{,} the outdated detectors are disposed of and replaced by new ones. The use of micro-detectors enables adaptability by design rather than attempting to change machine learning models and algorithms to support adaptability.

\subsection{Analyst Friendly Interpretation} 
 Adopting sophisticated machine learning techniques for malware detection in a production environment is a challenge. This is because most of the time it is not possible to understand how the machine learning systems make their malware detection decisions.  Therefore, tuning and maintaining these systems is a challenge in production and new techniques  for malware analysts to interpret and evaluate the performance of malware detectors are needed. We propose the use of evolutionary computation techniques such as genetic algorithms or clonal selection algorithms to generate an interpretation for black-box machine learning models such as deep learning. Using evolutionary computation, we could describe the decisions of malware detectors using a set of IF-Then rules. The only information required is the input features the malware detector uses to make a decision. 
 
 The IF-Then rules are useful to explain the behaviors that trigger a specific decision (e.g., malicious or benign) by the malware detector.  Cybersecurity and malware analyst are comfortable working with IF-Then rules. These rules will help in understanding the decision made by malware detectors, explain the scope of the detection, and identify potential over generalization or overfitting that could result in false positives or false negatives.
 
 It is essential that the IF-Then rules set interpretation of the malware detector to be expressed in raw malware behaviors and not in machine learning features. Machine learning features are most likely understandable by machine learning engineers and experts. The interpretation should be acceptable to a malware analyst who does not need to be machine learning experts. 

\subsection{Anti Adversarial Malware} 
 
 To improve the resilience of malware detectors against adversarial malware\textcolor{black}{,} we believe it is essential to study the effort required by the malware authors to design an adversarial malware for specific malware detectors. For example\textcolor{black}{,} what technique a malware author would use to probe and study a malware detector in production to design a malware that could bypass this detector. 

Measuring the effort to probe detectors and design adversarial malware under two main settings is essential. The first setting is black-box\textcolor{red}{,} where the malware authors have minimum knowledge about the malware detector internal design and the features used by the machine learning algorithm. The second setting is white-box\textcolor{black}{,} where the malware authors have sufficient knowledge about the malware detector internal design and the machine learning algorithm.   Training and updating the malware detectors is likely the most efficient solution against adversarial malware.  Knowing the effort needed to evade a malware detector will help in designing training strategies and policies to increase the effort required to evade the detectors.  

As we mentioned before\textcolor{black}{,} Cohen provided a formal proof that creating a perfect malware detection system is not possible \cite{cohen87,cohen89}. We believe that designing a perfect adversarial malware is not possible. Therefore we expect that using ensemble-based hybrid machine learning approach for malware detector will be effective against adversarial malware. It is expected that by creating a malware detector using an ensemble hybrid machine-learning approach, the risk of evading detection will decrease and the effort to design adversarial malware will increase. A hybrid machine learning model is when two or more different machine learning algorithms are used to construct the model. In the literature, adversarial malware samples evade malware detectors that use a single machine learning algorithm or technique \cite{Yang2017MalwareDI,Grosse17,Kolosnjaji18}. In our method, a hybrid machine learning approach for building a malware detector is an approach to provide a defense-in-depth model for malware detectors.

\section{\uppercase{Conclusion}}
\noindent In this paper, we reviewed the current state-of-the-art in malware detection using machine learning. We discussed the recent trend\textcolor{black}{s} in malware development and emerging malware threats. We argued that behavior\textcolor{black}{al} analysis would dominate the next generation antimalware systems. We discussed the challenges of applying machine learning to detect malware in the wild and proposed our thoughts on how we could overcome these challenges. Machine learning malware detectors require inexpensive training methods; they need to be interpretable for the malware analysts and not only for machine learning experts. Finally, they need to tolerate adversarial malware by design.  

\bibliographystyle{apalike}
{\small
\bibliography{example}}



\end{document}